\documentclass[letterpaper,12pt]{article}
\pdfoutput=1 
\usepackage{epsfig}
\usepackage{float}
\usepackage{dsfont}
\usepackage{jheppub} 

\usepackage{bbold}

\makeatletter
\def\@fpheader{\relax}
\makeatother

\usepackage{caption}
\usepackage{subcaption}

\usepackage{amsmath}
\usepackage{amssymb}
\usepackage{graphicx}

\renewcommand{\thanks}[1]{\footnote{#1}} % Use this for footnotes

\newcommand{\be}{\begin{equation}}
\newcommand{\bea}{\begin{eqnarray}}
\newcommand{\eea}{\end{eqnarray}}
\newcommand{\beq}{\begin{equation}}
\newcommand{\ee}{\end{equation}}

\def\ra{\rangle}

\def\simleq{\; \raise0.3ex\hbox{$<$\kern-0.75em
\raise-1.1ex\hbox{$\sim$}}\; }
\def\simgeq{\; \raise0.3ex\hbox{$>$\kern-0.75em
\raise-1.1ex\hbox{$\sim$}}\; }

\def\bi{\begin{itemize}}
\def\ei{\end{itemize}}

\def\sc{\setcounter{equation}{0}}

\usepackage{color}

\subheader{MIT-CTP/5029}

\title{Entanglement Holonomies}

\author[a]{Bart{\l}omiej Czech,} \author[b]{Lampros Lamprou,}\author[c]{Leonard Susskind}

\affiliation[a]{Institute for Advanced Study, Princeton, NJ 08540, USA}
\affiliation[b]{Center for Theoretical Physics, Massachusetts Institute of Technology, Cambridge, MA 02139-4307,USA}
\affiliation[c]{Stanford Institute for Theoretical Physics, Department of Physics, Stanford University\\
Stanford, CA 94305, USA\\[1cm]}

\abstract{We introduce a quantum notion of parallel transport between subsystems of a quantum state whose holonomies characterize the structure of entanglement. In AdS/CFT, entanglement holonomies are reflected in the bulk spacetime connection. When the subsystems are a pair of holographic CFTs in an entangled state, our quantum transport measures Wilson lines threading the dual wormhole. For subregions of a single CFT it is generated by the modular Berry connection and computes the effect of the AdS curvature on the transport of minimal surfaces. Our observation reveals a new aspect of the spacetime-entanglement duality and yet another concept shared between gravity and quantum mechanics.}

\begin{document}

\maketitle
\flushbottom
\tableofcontents

\section{Introduction}
Do quantum states have curvature? Certain quantum systems admit an effective description in terms of General Relativity in a higher dimensional AdS universe. The dynamical geometry of spacetime, in these examples, appears to represent properties of the system's state, e.g. the pattern of entanglement  \cite{markessay, Ryu:2006bv, Maldacena:2013xja}, its complexity \cite{Brown:2015bva} and others, suggesting a view of GR as another language for quantum mechanics. This idea was 
%BC recently 
summarized in \cite{Susskind:2017ney} as GR=QM and is supported by a number of similarities between familiar gravitational and quantum concepts. 

A central idea in General Relativity is that the orientations of local frames are related by a connection, whose holonomies reflect the curvature and topology of spacetime. We propose that quantum states exhibit a similar notion of parallel transport with non-trivial holonomies determined by the form of entanglement of their subsystems. Operationally, these holonomies are measured by transporting a probe state between subsystems via the quantum teleportation protocol, using the global state as resource.  

A concrete example of our proposal is the modular Berry connection, recently proposed in \cite{Czech:2017zfq}. It is introduced as a CFT connection that relates the frames of modular Hamiltonians of nearby subregions. In the bulk, modular Berry holonomies compute the precession of the local Rindler frame of a Ryu-Takayanagi surface, after a ``closed loop'' of deformations of its boundary support. This links our entanglement holonomies to the bulk spacetime connection, which we view as another non-trivial parallel in support of GR=QM and a step towards a precise articulation of a spacetime-entanglement duality.

In Section \ref{sec: 2}, we review the concept of quantum reference frames introduced in \cite{Aharonov:1967zza, Aharonov:1967zz} and explain how the form of entanglement determines their alignment.\footnote{For related interesting discussions see \cite{Bartlett:2006}.} In Section \ref{sec: 3}, we illustrate that quantum teleportation defines a transport between quantum frames that probes this alignment, a fact we then utilize in Section \ref{sec: 4} to argue that entanglement holonomies for a pair of holographic CFTs in the thermofield double state measure gravitational Wilson lines threading the dual wormhole. In Section \ref{sec: 5} we formulate bulk transport of Rindler frames in the CFT language and explain its quantum description in terms of the modular Berry connection.

\section{Aligned Quantum Frames}\label{sec: 2}

Reference frames are idealizations of physical systems such as clocks, meter sticks and gyroscopes, which we use as standards for comparison with our local measurements. When these reference systems are taken to be classical, General Relativity provides a geometric framework for relating their readings at separate locations. However, physical systems are quantum and so are our reference frames. In this section, we discuss the framework for describing alignment of quantum frames \cite{Aharonov:1967zza, Aharonov:1967zz}.

\begin{figure}[H]
\begin{center}
\includegraphics[scale=.35]{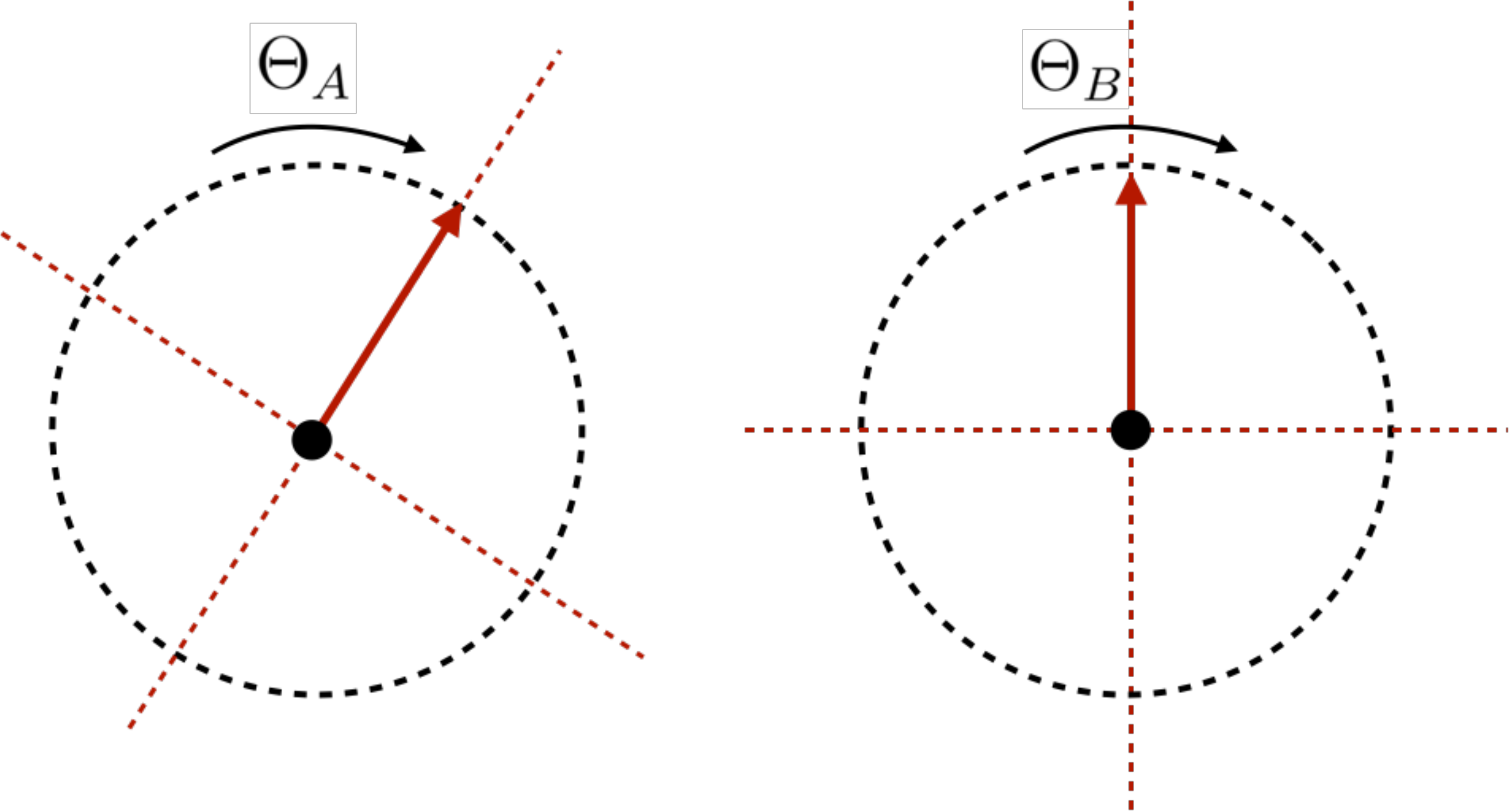}
\caption{\footnotesize{The orientation of a pair of gyroscopes can be used to define local Cartesian frames. Each of them is a quantum system described by the conjugate pair $(\Theta_i,L_i)$, $i=A,B$ which controls the direction in which they point.}}
\label{gyroscopes}
\end{center}
\end{figure}

Consider two gyroscopes at different locations in our lab and suppose for simplicity that their orientation is confined on a 2-D plane (fig.~\ref{gyroscopes}). Quantum mechanically, each gyroscope is described by a conjugate pair of operators $(\Theta_i,L_i)$, $i=A,B$ where $\Theta_i$ is the orientation on the plane and $L_i$ the corresponding angular momentum.

We can use our quantum gyroscopes to define Cartesian frames in their local neighborhoods. The relative orientation of the two frames can be made meaningful in two ways. The first is to simply transport gyroscope $A$ to the neighborhood of $B$ through the lab, align their orientation and then return $A$ to its original location. If gravity is unimportant, this parallel transport is trivial and would result in a state:
\begin{equation}
|AB\rangle = |\Theta\rangle_A |\Theta\rangle_B
\end{equation} 

A second method for aligning the Cartesian frames is to entangle the gyroscopes. Imagine preparing them, for example, in a state with vanishing total angular momentum:
\begin{equation}
|AB\rangle= \sum_{L=-L_{\text{max}}}^{L_{\text{max}}} \psi(L)\,  |L\rangle_A |-\! L\rangle_B \label{alignment}
\end{equation}
From the perspective of the global wavefunction, the orientation of both gyroscopes  is now random, since each of them is described by a reduced density matrix which is diagonal in the angular momentum basis. 

The orientation of B \emph{relative} to A, however, depends sensitively on the form of entanglement, dictated by the wavefunction $\psi(L)$. If, for example, we prepare the state $\psi(L) \propto e^{iL\theta}$ then, after a change of basis, (\ref{alignment}) becomes:
\begin{equation}
|AB\rangle = |L_{\text{tot}}=0\rangle |\Theta_{AB}=\theta\rangle
\end{equation}
where $\Theta_{AB}=\Theta_B-\Theta_A$. Therefore, according to A, gyroscope B points in a definite direction, at a relative angle of $\theta$. 

The definite relative alignment of the previous state should be contrasted with the case $\psi(L)=\delta_{L,0}$:
\begin{equation}
|AB\rangle = |0\rangle_A|0\rangle_B \label{uncorrelated}
\end{equation}
The relative orientation of A and B in this example has maximal quantum uncertainty and their corresponding Cartesian frames are uncorrelated.

The lesson of this discussion is that the alignment of quantum reference frames, which can be either definite or indefinite, is determined by the particular pattern of entanglement in the global state.

\section{Entanglement Holonomies}\label{sec: 3}

We now want to define a quantum notion of parallel transport. Given an entangled state of our quantum reference frames, we wish to devise a rule for transporting a probe state between them so that it ``learns'' about their relative orientation. Since the latter depends on the form of entanglement, quantum teleportation provides an appropriate rule, as we now illustrate.

In order to emphasize that our transport is quantum mechanical in nature and it does not rely on assumptions of robustness or classicality of the reference frames, we will consider the simplest type of gyroscopes: single qubits. 

Standard teleportation assumes an entangled state between reference frames $A$ and $B$, which we choose to be maximally entangled in a state of zero $\sigma_z$ component:

\begin{equation}
|AB \rangle =\frac{1}{\sqrt{2}}\left( e^{-i\frac{\theta}{2}}|u\rangle_A |d\rangle_B - e^{i\frac{\theta}{2}} |d\rangle_A |u\rangle_B\, \right).
\label{refstate}
\end{equation}
Here $u, d$ refer to spin along the $z$ axis and $\theta \in [0,2\pi)$ is an arbitrary phase. 

Consider a third system $C$ whose state $|\phi\rangle_C$ we will teleport from $B$ to $A$. The protocol proceeds with a joint measurement on $BC$ in the Bell basis. For simplicity, we will consider probabilistic teleportation so that if $BC$ is found in the singlet state

\begin{equation}
|\text{Bell}  \rangle_{BC} = \frac{1}{\sqrt{2}}\left(|u\rangle_B |d\rangle_C - |d\rangle_B |u\rangle_C\right)\,,
\end{equation}
then teleportation is successful and the state of $A$ gets updated with information about $C$. Otherwise we discard the state and repeat the protocol. 

At the end of the protocol, the state of $A$ is:
\begin{equation}
|\tilde{\phi}\rangle_A=\,_{BC}\langle \text{Bell}|\big(|AB\rangle \otimes |\phi\rangle_C\big)= e^{\frac{i}{2}\sigma_z \theta}|\phi\rangle_A \label{transportedC}
\end{equation}
which is rotated by an angle $\theta$ on the $xy-$plane with respect to $|\phi\rangle_C$. We can then return the teleported qubit to point $B$ by simple geometric transport through the lab and compare the result with a record of the original state $|\phi\rangle_C$. The comparison can be implemented experimentally by interfering $|\tilde{\phi}\rangle_A$ and $|\phi\rangle_C$ and eq. (\ref{transportedC}) will relate the observed fringe pattern to the relative phase in our resource state (\ref{refstate}). Similar statements hold for higher angular momentum gyroscopes, like those in Section~\ref{sec: 2}.

\begin{figure}[t]
\begin{center}
\includegraphics[scale=.35]{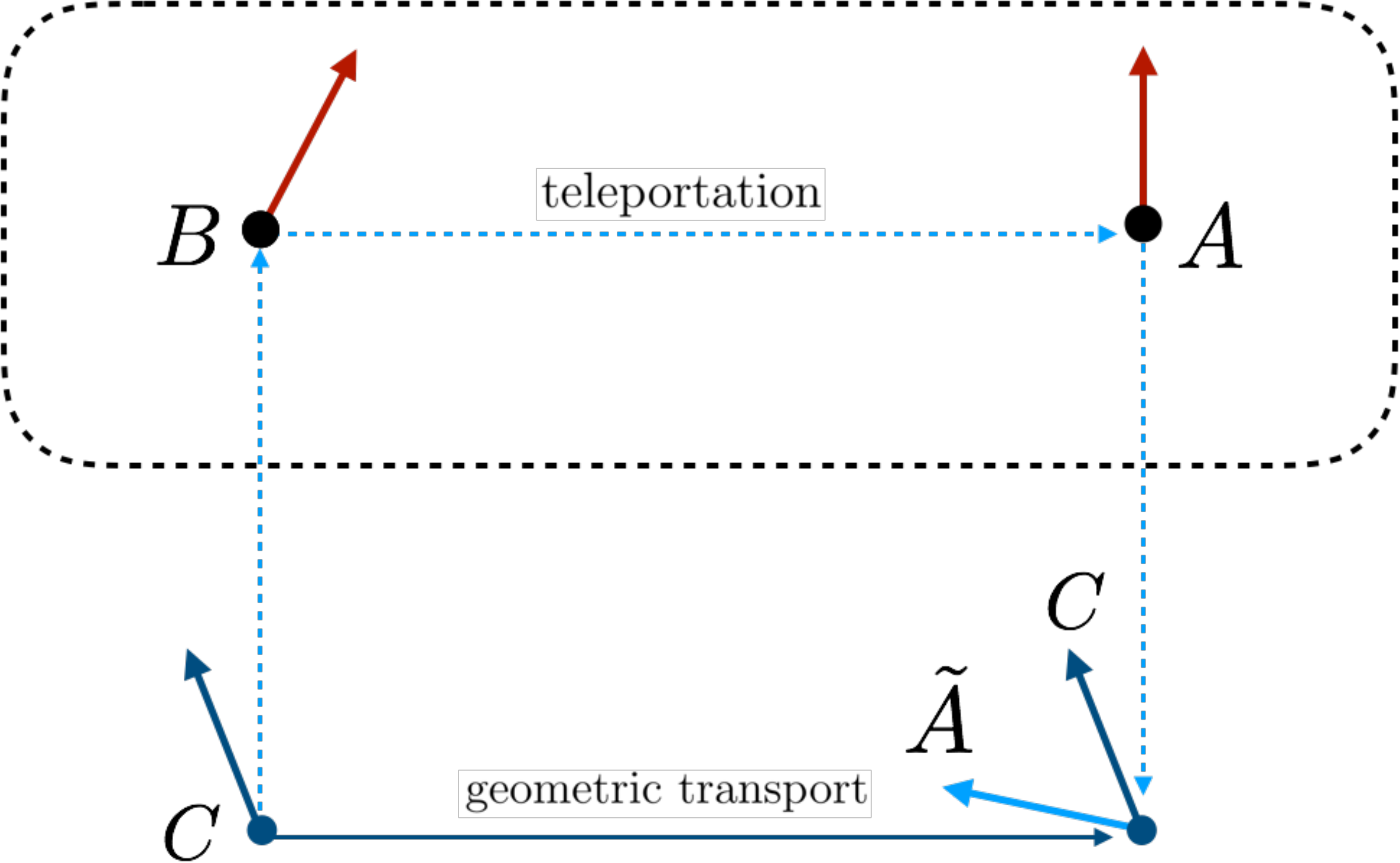}
\caption{\footnotesize{Transporting a probe gyroscope (C) around a closed loop. The solid blue line corresponds to standard spacetime transport of the gyroscope in the lab. The dashed light blue lines represent quantum transport of the gyroscope from reference frame B to A via quantum teleportation with reference state $|AB\rangle$ (\ref{refstate}). The angle between the gyroscopes transported along the two different paths measures the holonomy of the loop. }}
\label{transport}
\end{center}
\end{figure}

We interpret this procedure as the \emph{parallel transport} of the state of $C$ along a \emph{closed loop} (fig.~\ref{transport}). One leg of this path is ordinary geometric transport of the gyroscope through the lab while the other is quantum ``transport through the entangled state $|AB\rangle$". From the perspective of the ER=EPR proposal, we may interpret the latter as the parallel transport through a small Einstein-Rosen bridge. The relative orientation of the original and the transported state defines a quantum holonomy, induced by the form of entanglement of the reference state.

\section{Wormhole-Threading Wilson lines} \label{sec: 4}

We argue that entanglement holonomies, as defined in the previous section, translate to geometric holonomies in spacetime when applied to CFTs with a holographic dual. We can substitute  the two gyroscopes of our toy example with a pair of large $N$ CFTs that Alice and Bob have engineered on two shells of matter in their lab. Moreover, they have entangled them in the thermofield double state so that they admit a dual description in terms of a two-sided AdS black hole. 

\paragraph{Rotation.} Alice and Bob can choose to further entangle the angular momenta of their CFTs with relative phases as in (\ref{refstate}):
\begin{equation}
|\psi\ra  = \sum_{L,E} e^{-\beta E/2} \, e^{iL\theta}\, |L,E\ra \otimes |\!-\! L,E\ra \label{TFDtheta}
\end{equation}

The angle $\theta$ relates the orientation at the two ends of the wormhole. As before, the experimentalists can measure it by first aligning two gyroscopes in the lab using ordinary transport and then ``closing a loop'' with Alice teleporting her gyroscope's state to Bob using (\ref{TFDtheta}). 

According to recent results \cite{Bennett, Gao:2016bin, Maldacena:2017axo, Susskind:2014yaa}, however, the quantum teleportation protocol on the boundary renders the bulk wormhole temporarily traversable, allowing the teleported system to physically propagate to the other side through the open bridge. Therefore, when the quantum systems are holographic
the quantum transport we defined in the previous section becomes geometric transport through the wormhole. The holonomy Alice and Bob measure when they read off the misalignment of the transported gyroscopes is, from the bulk perspective, induced by a gravitational Wilson line threading the wormhole.

\paragraph{Time translation.} We could also imagine that Alice and Bob prepared their system in a state with some energy-dependent relative phases:
\begin{equation}
|\psi'\ra  = \sum_{L,E} e^{-\beta E/2}\, e^{iE\eta}\, |L,E\ra \otimes  |\!-\! L,E\ra \label{TFDeta}
\end{equation}
This could be done by applying some amount of time evolution on the two sides independently. In this case, the phases determine the relative alignment of the Schwarzchild clocks on the two sides of the wormhole.  The holonomy induced by this entanglement pattern can again be measured in the lab. Alice and Bob can sync a pair of clocks and use them to define time on each of their CFTs. When Alice teleports her clock to Bob, he will observe that the two clocks' readings are off by a time-shift equal to $\eta$. A similar setup was recently discussed in \cite{vanBreukelen:2017dul}.

For a bulk observer who traverses the wormhole opened via the Gao-Jafferis-Wall method, this time-shift induced by the entanglement of (\ref{TFDeta}) is an extra contribution to his time dilation with respect to boundary time, in addition to the familiar effect stemming from spacetime curvature. The internal observer attributes this, once more, to the non-trivial spacetime connection induced by a wormhole-threading Wilson line.

\newpage
\section{Transport in AdS and Modular Berry Connection}\label{sec: 5}
Spacetime holonomies are also induced by the curvature of spacetime. When curvature is non-vanishing, parallel transport of an observer's local Lorentz frame around a closed loop will result in a Lorentz transformation.  In AdS/CFT this gravitational phenomenon has its origin in the entanglement pattern between subregions in a single CFT and it is captured by the modular Berry connection \cite{Czech:2017zfq, Czech:upcoming}, as we now explain.

\subsection{Geometric Transport of Minimal Surfaces}

\begin{figure}[t!]
\begin{center}
\includegraphics[scale=.35]{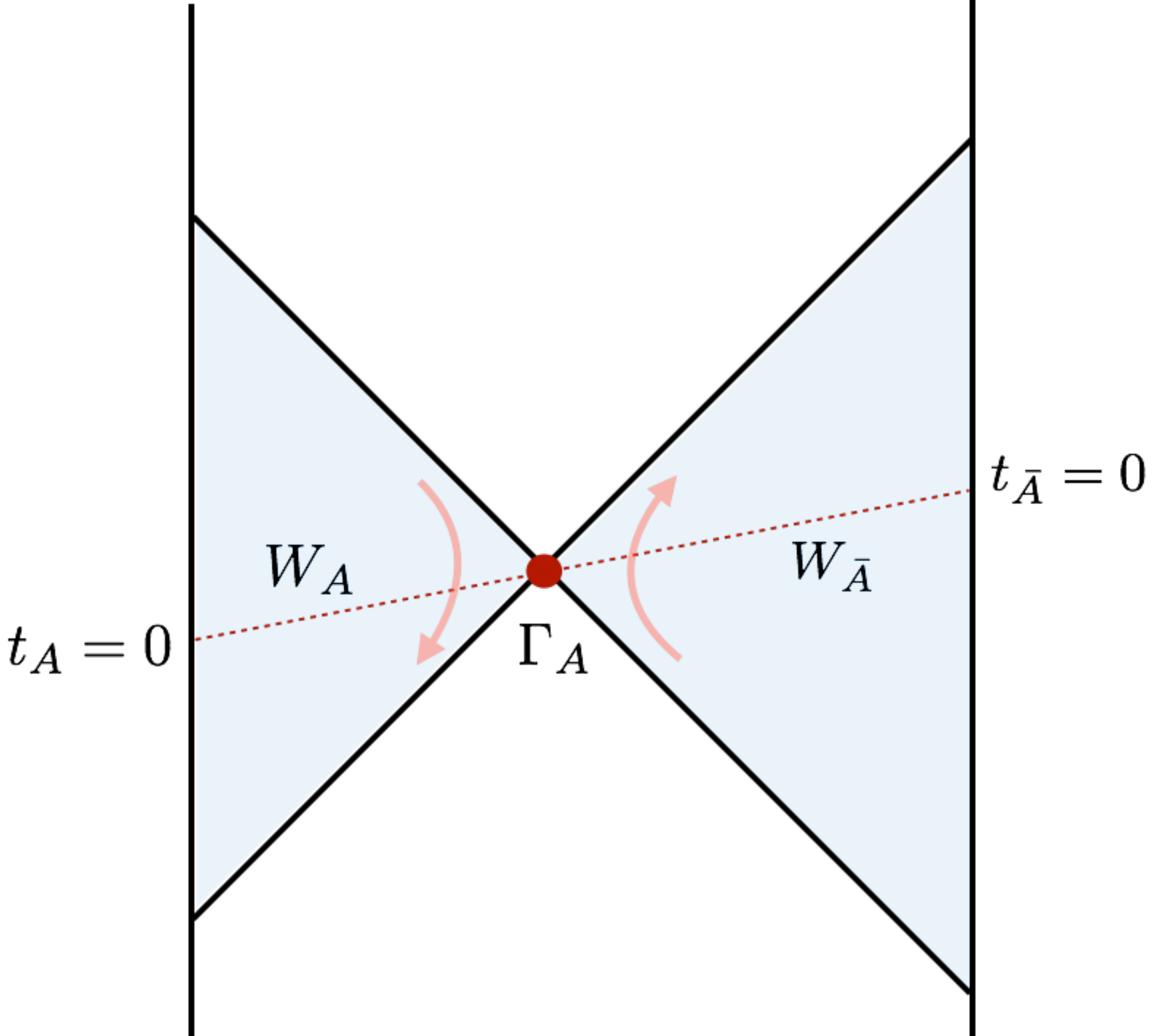}
\caption{\footnotesize{A minimal surface $\Gamma_A$ in AdS and the associated pair of Rindler wedges $W_A, W_{\bar{A}}$. The red arrows represent Rindler time translations in the two wedges. The combined transformation $(t_A, t_{\bar{A}})\rightarrow (t_A+\epsilon, t_{\bar{A}}-\epsilon)$ is an isometry of AdS and acts as a boost in the vicinity of $\Gamma_A$. Fixing the Rindler frame of $\Gamma_A$ amounts to choosing an origin for the Rindler clocks. The clocks on $W_A$, $W_{\bar{A}}$ are synced, as dictated by the vacuum state (\ref{vacuum}). }}
\label{rindlerframe}
\end{center}
\end{figure}

We will sketch the idea in pure AdS; a more complete and rigorous discussion will appear in \cite{Czech:upcoming}. Moreover, instead of transporting the local frame of a point along a curve in spacetime, we will consider the transport of a codimension-2 minimal surface. A minimal surface $\Gamma_A$ partitions spacetime in two complementary Rindler wedges $W_A$, $W_{\bar{A}}$. The bulk metric is invariant under a coordinated time translation in both wedges $(t_A, t_{\bar{A}})\rightarrow (t_A+\epsilon, t_{\bar{A}}-\epsilon)$ (fig.~\ref{rindlerframe}). These transformations preserve the location of the surface and act as Lorentz boosts in its vicinity.

In order to describe the bulk from the perspective of the minimal surface we need to select a Rindler frame. Physically, this amounts to choosing  a Rindler clock for the wedges or, equivalently, a boost angle coordinate in the neighborhood of the horizon. The Rindler frame of the minimal surface is, therefore, the analogue of the Lorentz frame on the tangent space of a point in the ordinary discussion of parallel transport.

We can now transport $\Gamma_A$ to another infinitesimally separated surface and study the transformation of the associated Rindler frame. While the general rules for generating this transport need to be carefully defined---and will be the subject of \cite{Czech:upcoming}---in the vacuum we can move in the space of minimal surfaces by simply utilizing the AdS isometries. For example, transport around the infinitesimal loop $\Gamma_A\rightarrow \Gamma_B\rightarrow \Gamma_C \rightarrow \Gamma_A$ of fig.~\ref{rindlertransport} will induce a holonomy
\begin{equation}
U_{AC}U_{CB}U_{BA}=S_A, \label{conformalloop}
\end{equation}
where $U_{ji}$ are the isometries that map $\Gamma_i$ to $\Gamma_j$ and $S_A\in SO(1,d-1)\times SO(1,1)$ belongs to the stabilizer group of the minimal surface $\Gamma_A$. The $SO(1,d-1)$ component of the holonomy computes an overall translation and rotation of the internal coordinates on the minimal surface and was discussed in detail in \cite{Czech:2017zfq}. The $SO(1,1)$ component, on the other hand, measures the Rindler boost of the frame of the minimal surface. In what follows we focus on this boost component of the holonomy. 

The origin of the holonomy is the curvature of AdS. The isometries $U_{ji}$ in (\ref{conformalloop}) map individual points on $\Gamma_i$ and their local tangent spaces to corresponding points on $\Gamma_j$ via ordinary parallel transport. Upon closing a loop, every point on $\Gamma_A$ gets parallel transported to some new point on it by following a curve in AdS and its local Lorentz frame can be compared with the original frame at that location. The difference between the two measures a spacetime holonomy.  In a general geometry this boost angle may vary along the minimal surface, but in pure AdS the homogeneity of the space-time reduces it to a global shift of the Rindler clock. 

In a nutshell, spacetime geometry relates the internal clocks of different Rindler wedges via an integrated version of the bulk connection that parallel transports their horizons.

\subsection{Modular Berry Holonomies}
The two wedges $W_A$, $W_{\bar{A}}$ selected by the minimal surface $\Gamma_A$ are described by the reduced density matrices of their dual boundary subregions, $\rho_A, \,\rho_{\bar{A}}$ respectively. Their modular Hamiltonians, defined by $H_{mod}=-\log \rho$, are Hermitian CFT operators that generate Rindler time translations in the corresponding bulk wedge. The choice of a Rindler clock in $W_A$ is, therefore, mapped to the selection of an origin for modular time $(t=0)$ in subregion $A$ or, equivalently, to a convention for the phases of the modular eigenstates $e^{iEt}|E\rangle_A$.

In the modular eigenbasis, the global vacuum state reads:
\begin{equation}
|0\rangle_{CFT}= \frac{1}{Z}\sum_E e^{-\pi E} |E\rangle_A |E\rangle_{\bar{A}}. \label{vacuum}
\end{equation}
Recalling our discussion of eq.~(\ref{TFDeta}), the absence of relative phases in (\ref{vacuum}) implies that the Rindler clocks on the opposite sides of the minimal surface are synced (fig.~\ref{rindlerframe}). Furthermore, it is evident from (\ref{vacuum}) that the state is annihilated by the full modular Hamiltonian:
\begin{equation}
\left(H_{\text{mod},A}-H_{\text{mod},\bar{A}}\right)|0\rangle_{CFT}=0 \label{envariance}
\end{equation}
This is the boundary incarnation of the Rindler boost symmetry that preserves the minimal surface in AdS. For every CFT bipartition, there is a ``local'' gauge ambiguity in the frame of the \emph{full modular Hamiltonian}. This ambiguity is the boundary avatar of the bulk freedom to pick a Rindler frame for $\Gamma_A$ (fig.~\ref{rindlerframe}).

\begin{figure}[t!]
\centering
\begin{tabular}{ccc}
\includegraphics[width=.48\textwidth]{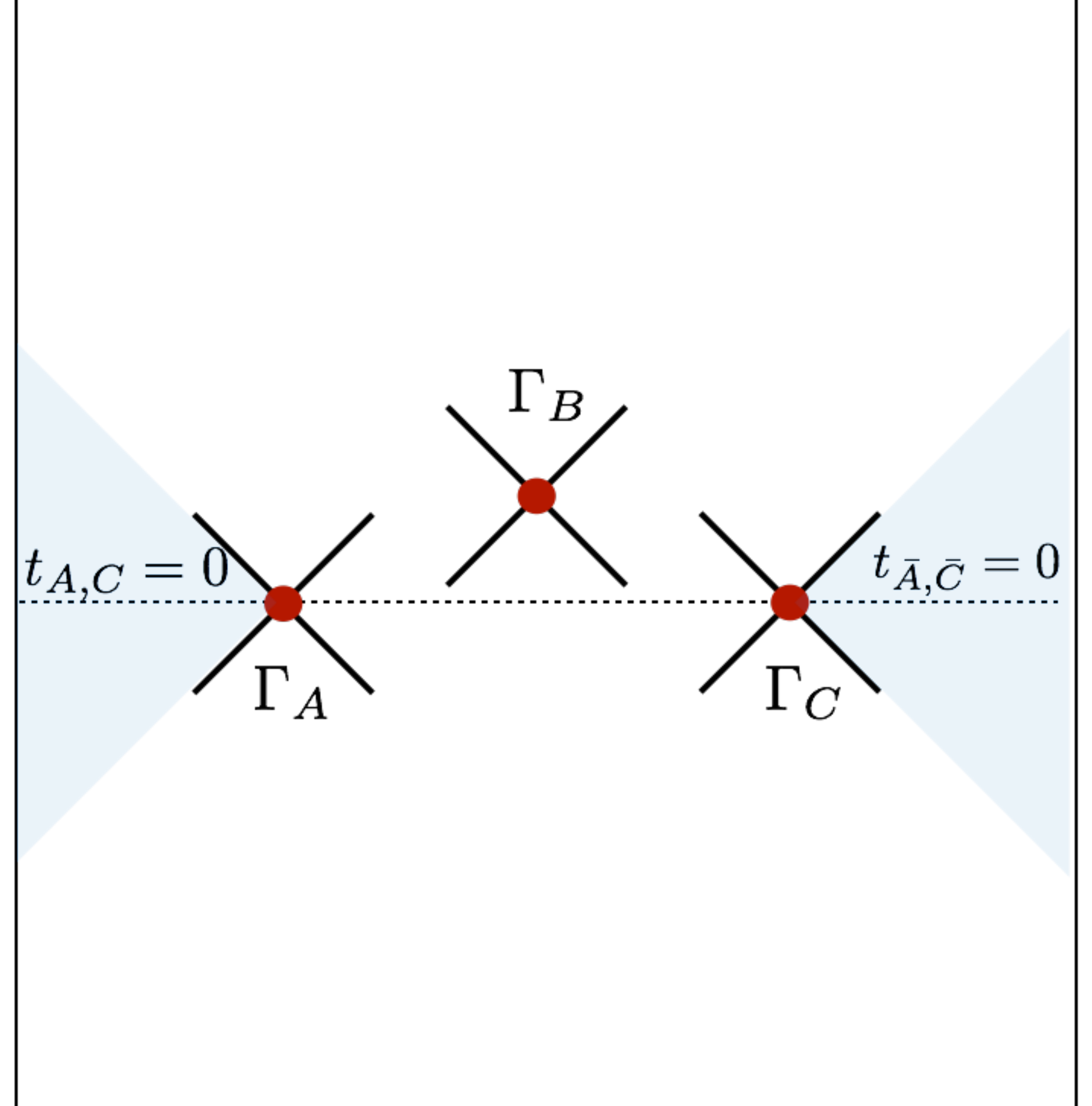} & 
\includegraphics[width=.48\textwidth]{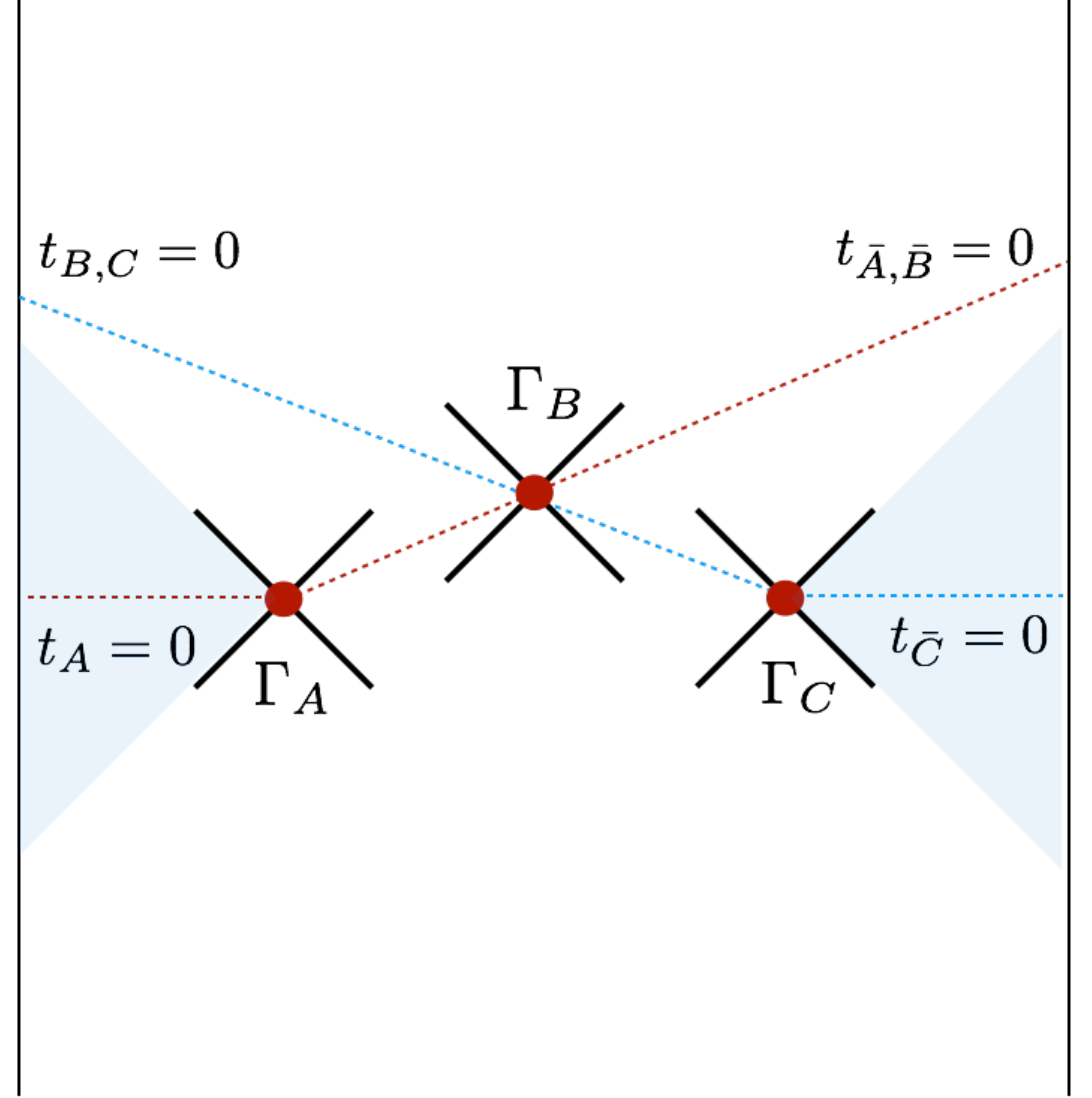} 
\end{tabular}
\caption{\footnotesize{Three infinitesimally minimal surfaces in AdS, $\Gamma_A,\Gamma_B,\Gamma_C$, can be used to close a loop of modular Hamiltonians and compute the precession of the modular frame. The first leg of the closed path is transport of a Rindler clock directly from $A$ to $\bar{C}$ (LEFT). The invariance of the state under (\ref{envariance}) can be utilized to sync the Rindler clocks ($t_A=t_C$ and $t_{\bar{A}}=t_{\bar{C}}$) and trivialize the transport. The second leg is transport from $A$ to $\bar{B}$ and then to $\bar{C}$ (RIGHT). The clock of $\bar{B}$ is now misaligned with both $A$ and $\bar{C}$ as can be seen by the kinks in the red and blue AdS time-slices. The modular Berry holonomy (\ref{BerryHolonomy}) depends on the hyperbolic angles of these kinks.} }
\label{rindlertransport}
\end{figure}

\paragraph{Transport of modular frames.} Our task, now, is to transport a Rindler clock from a subregion to another in two different ways in order to form a loop and measure a spacetime holonomy. The closed loop we will consider is composed by three infinitesimal steps, passing through the minimal surfaces $\Gamma_A$, $\Gamma_B$ and $\Gamma_C$ illustrated in fig.~\ref{rindlertransport}. According to our discussion in Sec.~\ref{sec: 2}, the result of the transport in the CFT should be reflected in the form of entanglement between the modular eigenbases of the subregions along our path.

We can exploit the modular gauge freedom (\ref{envariance}) to align the modular clocks of $(A,\bar{A})$ and $(C,\bar{C})$, simply by synchronizing them with the AdS time-slice that contains both minimal surfaces (fig.~\ref{rindlertransport}). In this convenient gauge, teleportation of a clock from $A$ to $\bar{C}$ using (\ref{vacuum}) will result in no time-translation. To be more precise, when $\Gamma_A$ and $\Gamma_C$ are infinitesimally separated, we can think of $|E\rangle_{\bar{A}}$ and $|E\rangle_{\bar{C}}$ as states in the same Hilbert space \cite{Faulkner:2016mzt}. A convenient gauge choice for the modular frames sets the phases of the $H_{\text{mod},\bar{A}}$ and $H_{\text{mod},\bar{C}}$ eigenvectors so that $\, _{\bar{C}}\langle E|E\rangle_{\bar{A}}$ is real. In this gauge, the state of $A\bar{C}$, which in the limit $\bar{C}\rightarrow\bar{A}$ is approximately given by
\begin{equation}
|A\bar{C}\rangle\approx \frac{1}{Z}\sum_E e^{-\pi E}\, _{\bar{C}}\langle E|E\rangle_{\bar{A}} \,|E\rangle_A |E\rangle_{\bar{C}}, \label{ACstate1}
\end{equation}
comes with no relative phases between $A$ and $\bar{C}$ and the two clocks are synced.

On the other hand, the alignment of $A$ and $\bar{B}$ in the above gauge is non-trivial (fig.~\ref{rindlertransport}). If we set the origin of modular time for $(B,\bar{B})$ to be the Rindler time-slice that contains both $\Gamma_B$ and $\Gamma_A$ then $\, _{\bar{B}}\langle E|E\rangle_{\bar{A}}=c_E \,e^{iE\eta_{A}}$, where $c_E\in \mathbb{R}$ and $\eta_A$ is the hyperbolic angle of the ``kink'' of the red-colored time-slice at the minimal surface $\Gamma_A$ in fig.~\ref{rindlertransport}. As a result, the state of $A\bar{B}$ is approximately:
\begin{align}
|A\bar{B}\rangle &\approx \frac{1}{Z}\sum_E e^{-\pi E}\, _{\bar{B}}\langle E|E\rangle_{\bar{A}} \,|E\rangle_A |E\rangle_{\bar{B}} \nonumber\\
&\approx \frac{1}{Z}\sum_E e^{-\pi E}\,c_E \,e^{iE\eta_{A}} \,|E\rangle_A |E\rangle_{\bar{B}} \label{ABstate}
\end{align}
The two Rindler clocks are misaligned by $\eta_{A}$. 

To close the loop, we need to relate the frames of the $H_{\bar{B}}$ and $H_{\bar{C}}$ eigenvectors. As before, this is determined by the inner product $\,_{\bar{C}}\langle E|E\rangle_{\bar{B}}$ which allows us to obtain a state for $A\bar{C}$ from (\ref{ABstate}):
\begin{align}
|\widetilde{A\bar{C}}\rangle &\approx \frac{1}{Z}\sum_E e^{-\pi E} \, _{\bar{C}}\langle E|E\rangle_{\bar{B}} \, _{\bar{B}}\langle E|E\rangle_{\bar{A}} \,|E\rangle_A |E\rangle_{\bar{C}}  \label{ACstate2}
\end{align}
As fig.~\ref{rindlertransport} illustrates, with the above gauge choices $\,_{\bar{C}}\langle E|E\rangle_{\bar{B}}$ is imaginary with phase equal to $E(\eta_{C}-\eta_B)$, where $\eta_{B},\eta_C$ are the boost angles of the ``kinks'' at $\Gamma_B$ and $\Gamma_C$, respectively. This is because in order to make $\,_{\bar{C}}\langle E|E\rangle_{\bar{B}}$ real, namely to sync the clocks of $\bar{B}$ and $\bar{C}$, we would need to boost the state of $\bar{B}$ by $\exp\left[{iH_{\text{mod},\bar{B}}\,\eta_B}\right]$ and the state of $\bar{C}$ by $\exp\left[{iH_{\text{mod},\bar{C}}\,\eta_C}\right]$.

\paragraph{Modular Berry holonomy and bulk curvature.} The relative phase between the modular eigenstates $|E\rangle_A$ and $|E\rangle_{\bar{C}}$ differs between the two ways of relating them---(\ref{ACstate1}) and (\ref{ACstate2})---by:
\begin{equation}
\text{Im} \log \left[\, _{\bar{A}}\langle E|E\rangle_{\bar{C}} \, _{\bar{C}}\langle E|E\rangle_{\bar{B}}\,_{\bar{B}}\langle E|E\rangle_{\bar{A}}\right]=E(\eta_B+\eta_A-\eta_C) \label{BerryHolonomy}
\end{equation}
This expression holds independently of our convenient gauge choice $\, _{\bar{C}}\langle E|E\rangle_{\bar{A}} \in \mathbb{R}$.

In the continuum limit the left-hand side computes the \emph{Berry curvature} \cite{berrycurv}, which arises from taking the modular Hamiltonian around an infinitesimal loop from $H_{\text{mod}, \bar{A}}$ to $H_{\text{mod}, \bar{B}}$ to $H_{\text{mod}, \bar{C}}$ and back to $H_{\text{mod}, \bar{A}}$. The right-hand side measures the boost angle by which the Rindler frame of the minimal surface precesses. It is non-zero due to the curvature of the AdS space-time. 

We have, therefore, arrived at a direct link between the \emph{geometric connection} in gravity and the \emph{Berry connection} in quantum mechanics. The Berry connection of interest here, however, is not the familiar one describing evolution under time-dependent Hamiltonians, but instead a Berry connection relating the eigenspaces of modular Hamiltonians of different subsystems of the CFT. The modular Berry connection is the correct tool for studying the relative alignment of quantum frames, which originates from their entanglement properties in the global wavefunction.

A more rigorous treatment of the modular Berry connection for the CFT vacuum utilizes the geometric properties of the space of vacuum modular Hamiltonians called kinematic space \cite{Czech:2015qta, Czech:2016xec}; the details of this method were explained in \cite{Czech:2017zfq}. The extension of these ideas to other asymptotically AdS spacetimes and general CFT states will be presented in upcoming work \cite{Czech:upcoming}.

\section{Conclusion}\label{sec: 6}
When classical observers at separate locations measure an observable of a given system they generically disagree on the outcome. Part of the mismatch stems simply from their individual choice of orientation and grading of their measuring devices; part of it, however, is physical and can be extracted from the relations of a family of observers around a closed loop. The connection of General Relativity and its curvature encapsulate such relative alignments for measurements of distances, time shifts, rotations and boosts. 

Similarly, to meaningfully compare their observations of a quantum system, different experimentalists need to relate their Hilbert space bases. For a given observable $\mathcal{O}$, their conventions may differ by the action of any symmetry of $\mathcal{O}$, including the trivial transformation of rotating its eigenstates by a phase. Although this choice of ``quantum frame'' is inconsequential for experiments executed by a single observer, it is essential for determining the map between reference frames.

Our first central claim was that when the observers themselves---or their devices---are treated as internal quantum systems, their relative bases are determined by the way they are entangled in the global state. The observers can operationally probe this alignment by exchanging a probe state via quantum teleportation, which we interpreted as a quantum mechanical notion of transport with non-trivial holonomies. Quantum systems are, therefore, endowed with a connection encoding their entanglement structure. It would be interesting to explore a potential overlap of this connection with the emergent gauge field ideas of \cite{Harlow:2015lma}.

We, further, argued that our proposed quantum connection and the geometric connection in spacetime are identified under the AdS/CFT duality. This potentially provides a new diagnostic for the emergence of a classical bulk spacetime. The quantum holonomies computed by the modular Berry connection on the boundary are not expected to be geometric in general. Equation (\ref{BerryHolonomy}) produces a unitary rotation for every eigenspace of the modular Hamiltonian, with the transformations of different eigenspaces being generically uncorrelated. In the vacuum example of Section~\ref{sec: 5}, conformal symmetry ensured that the holonomies were linear in the modular eigenvalues. This allowed us to interpret them as some amount of modular flow and, therefore, a geometric boost of the frame of the bulk minimal surface. The requirement that the modular Berry connection is, at leading order in $N$, valued in the local symmetry group of a minimal surface---local boosts and surface diffeomorphisms---appears necessary for a boundary state to be consistent with a classical bulk.

The idea of using quantum teleportation to extract geometric information about the bulk is reminiscent of the discussion in \cite{Czech:2014tva}. In that work, the length of bulk curves in $AdS_3$, as computed by differential entropy \cite{diffent}, was identified with the cost of a communication protocol based on state merging \cite{merging}, a simple generalization of teleportation. But differential entropy was also shown in \cite{Czech:2017zfq} to equal a translational component of the modular holonomy. This suggests that the protocol of \cite{Czech:2014tva} may serve as a useful guide for generalizing the quantum transport proposed in Section~\ref{sec: 3} to other situations, where state merging and not just teleportation is involved.

\acknowledgments{BC and LL are grateful to Jan de Boer and Dongsheng Ge for collaboration on related ongoing work \cite{Czech:upcoming}. We also thank Gilad Lifschytz, Hong Liu, Sam McCandlish, Rob Myers, Xiao-Liang Qi, James Sully, Wati Taylor, Erik Verlinde, Herman Verlinde and Jieqiang Wu for useful discussions. BC and LL acknowledge the hospitality of MITP during the ``Modern Techniques for AdS and CFT'' program where these ideas were first presented, and of the Galileo Galilei Institute during the ``Entanglement in Quantum Systems'' program. The work of BC is supported by NSF grant PHY-1606531 and by Swarthmore College. LL is supported by the Pappalardo Fellowship. LS's research supported by NSF Award Number 1316699.}

\end{document}